\documentclass[apj]{emulateapj}

\pdfoutput=1

\usepackage{natbib} 
\usepackage{graphicx}
\usepackage{threeparttable}
\usepackage{amsmath}
\usepackage{color}
\usepackage{booktabs}
\usepackage{multirow} 
\usepackage{etex,xy} 
\xyoption{all} 
\usepackage{hyperref}

\definecolor{orange}{rgb}{0.93, 0.57, 0.13}
\definecolor{green}{rgb}{0.01, 0.75, 0.24}

\newcommand{\MIR}{\texttt{MIR}}

\newcommand{\CASA}{\texttt{CASA}}
\newcommand{\CLEAN}{\texttt{CLEAN}}
\newcommand{\vissample}{\texttt{vis-sample}}
\newcommand{\RADMC}{\texttt{RADMC-3D}}
\newcommand{\emcee}{\texttt{emcee}}

\newcommand{\ie}{\emph{i.e.}}
\newcommand{\eg}{e.g.}
\newcommand{\emm}[1]{\ensuremath{#1}}   
\newcommand{\emr}[1]{\emm{\mathrm{#1}}} 

\newcommand{\Tkin}{\emm{T_\emr{kin}}}

\newcommand{\Tex}{\emm{T_\emr{ex}}}

\newcommand{\Td}{\emm{T_\emr{dust}}}

\newcommand{\Msun}{\emr{M_\odot}}

\newcommand{\hh}{\emr{H_2}}                  
\newcommand{\hho}{\emr{H_2O}}                  

\newcommand{\hhco}{\emr{H_2CO}}                  
\newcommand{\phhco}{\emr{p-H_2CO}}                  
\newcommand{\ohhco}{\emr{o-H_2CO}}                  
\newcommand{\nhhh}{\emr{NH_3}}                  

\renewcommand{\deg}{\emm{^\circ}}
\newcommand{\pccm}{~\rm{cm}^{-3}}

\newcommand{\kms}{\emr{\,km\,s^{-1}}}

\newcommand{\Tspin}{\emm{T_\emr{spin}}}

\begin{document}


\title{H$_2$CO ortho-to-para ratio in the protoplanetary disk HD~163296} 


\author{V.V. Guzm\'an\altaffilmark{1}}
\email{viviana.guzman@alma.cl}

\author{K.I. \"Oberg\altaffilmark{2}}

\author{J. Carpenter\altaffilmark{1}}

\author{R. Le Gal\altaffilmark{2}}


\author{C. Qi\altaffilmark{2}}

\author{J. Pagues\altaffilmark{2}}

\altaffiltext{1}{Joint ALMA Observatory (JAO), Alonso de C\'ordova 3107 Vitacura, Santiago de Chile, Chile}
\altaffiltext{2}{Harvard-Smithsonian Center for Astrophysics, 60 Garden Street, Cambridge, MA 02138, USA}

\begin{abstract}
Ortho-to-para (o/p) ratios of species like water, ammonia and
formaldehyde (\hhco{}) are believed to encode information about the
formation history of the molecule. Measurements of o/p ratios in
protoplanetary disks could thus be used to constrain their physical
and chemical histories. We present the first measurement of the
\hhco{} o/p ratio in a protoplanetary disk, using three ortho and two
para lines observed with the Sub-millimeter Array (SMA) combined with
one highly resolved measurement of a single \hhco{} line with ALMA
toward the disk around Herbig Ae star HD~163296. We find a
disk-averaged \hhco{} o/p ratio of 1.8-2.8 (depending on the assumed
disk structure), corresponding to a spin temperature of $11-22$~K. We
also derive a rotational temperature of 24~K from the flux ratio of
the three ortho lines. The observed spatial distribution, as seen by
ALMA, as well as the rotational temperature and the o/p ratio, at the
large scales the SMA is most sensitive to, are consistent with a
low-temperature formation pathway, most likely grain surface
chemistry, of \hhco{} in this disk.
\end{abstract}






 \newcommand{\TabObs}{%
   \begin{table*}
     \begin{center}
       \caption{Observational parameters.} 
       \label{tab:obs}
       \begin{tabular}{ccccccc}\toprule
         Line & Freq.   & Chan width & Beam      & PA & Chan. rms & Mom 0 rms$^a$ \\
         & GHz & \kms{} & '' & $\deg$ & mJy beam$^{-1}$ & mJy beam$^{-1}$ \kms{} \\
         \midrule
         \multicolumn{7}{c}{\it SMA observations}\\
         \phhco{} $3_{03}-2_{02}$  & 218.222 &  0.6  & $6.91\times2.12$ & 32.36 & 50.93  & 99.85  \\
         \ohhco{} $3_{13}-2_{12}$  & 225.698 &  0.6  & $3.38\times2.12$ & 56.04 & 53.26  & 105.67 \\
         \ohhco{} $4_{14}-3_{13}$  & 281.527 &  0.6  & $4.61\times3.22$ & 12.76 & 80.74  & 145.35 \\
         \phhco{} $4_{04}-3_{03}$  & 290.623 &  0.6  & $7.68\times4.75$ & 45.29 & 112.05 & 280.79 \\
         \ohhco{} $4_{13}-3_{12}$  & 300.837 &  0.6  & $7.24\times4.60$ & 45.17 & 118.25 & 271.34 \\
         \multicolumn{7}{c}{\it ALMA observations}\\                                               
         \phhco{} $4_{04}-3_{03}$  & 290.623 & 0.6  & $1.00\times0.90$ & -72.80 & 7.22  &  15.90 \\
         \bottomrule
       \end{tabular}
     \end{center}
     $^a$The cubes are integrated from 2 to 10\kms{}.\\
   \end{table*}
 }
 
 \newcommand{\TabFluxes}{%
   \begin{table}
     \begin{center}
       \caption{Upper energies, critical densities and integrated fluxes.} 
       \label{tab:fluxes}
       \begin{tabular}{cccc}\toprule
         Line & $E_u$ & $n_{crit}$ & Flux density\\
         & K & $\pccm$ & Jy~\kms{} \\
         \midrule
         \multicolumn{4}{c}{\it SMA observations}\\         
         \phhco{} $3_{03}-2_{02}$  & 21.0 & $2.6\times10^6$ & $0.89\pm0.28$\\
         \ohhco{} $3_{13}-2_{12}$  & 33.4 & $4.3\times10^6$ & $0.98\pm0.22$\\
         \ohhco{} $4_{14}-3_{13}$  & 45.6 & $7.5\times10^6$ & $1.93\pm0.28$\\
         \phhco{} $4_{04}-3_{03}$  & 34.9 & $5.8\times10^6$ & $1.31\pm0.11$\\
         \ohhco{} $4_{13}-3_{12}$  & 47.9 & $8.8\times10^6$ & $1.61\pm0.30$\\
         \multicolumn{4}{c}{\it ALMA observations}\\                  
         \phhco{} $4_{04}-3_{03}$  & 34.9 & $5.8\times10^6$ & $1.22\pm0.05$\\
         \bottomrule
       \end{tabular}
     \end{center}
   \end{table}
 }

 \newcommand{\TabModelParam}{%
   \begin{table*}[t!]
     \begin{center}
       \caption{Model parameters.}
       \label{tab:param}
       \begin{tabular}{ll|llll}\toprule
         \multicolumn{2}{c|}{\it Fixed} & \multicolumn{4}{c}{\it Best fit parameters}\\
         \multicolumn{2}{c|}{\it parameters} & & $(z/r)_l=0.15$ & $(z/r)_l=0.20$ & $(z/r)_l=0.25$ \\
         \midrule
          & &  \multicolumn{4}{c}{\it ALMA data}\\
         $R_1$ & 1~au & $X_1$      & $(3.7\pm0.6) \times 10^{-10}$ & $(2.7\pm0.5) \times 10^{-9}$ & $(1.2\pm0.2) \times 10^{-8}$\\
         $R_2$ & 80~au  & $X_2$    & $(1.2\pm0.1) \times 10^{-10}$ & $(7.6\pm0.6) \times 10^{-10}$ & $(5.3\pm0.5) \times 10^{-9}$ \\
         $R_{out}$ & 600~au      & $\alpha$ & $0.52\pm0.09$ & $-0.34\pm0.10$ & $-1.23\pm0.12$\\
         $(z/r)_u$  & 0.5    &  $\chi^2_{red}$  & 1.0716999 & 1.0717228 & 1.0717574 \\  
         \cmidrule{3-6}
         & & \multicolumn{4}{c}{\it SMA data}\\
         & & $o/p$   & $2.50^{+0.31}_{-0.27}$ & $2.23^{0.28}_{-0.24}$ & $2.00^{+0.26}_{-0.22}$ \\
         & & $scale$ & $1.17\pm0.06$ & $1.20\pm0.06$ & $1.32\pm0.06$ \\
         & & $\chi^2_{red}$ &    1.0479973 & 1.0479993 & 1.0480178\\
         \bottomrule
       \end{tabular}
     \end{center}
   \end{table*}
 }

 \newcommand{\TabOPRs}{%
   \begin{table}[t!]
     \begin{center}
       \caption{Measured o/p ratios in different environments.}
       \label{tab:oprs}
       \begin{tabular}{lcccc}\toprule
         Source & \nhhh{} o/p & \hho{} o/p & \hhco{} & Refs.\\
         \midrule
         \multicolumn{5}{l}{\it Comets} \\
         73P-B    & 1.01$\pm$0.03 & 3.20$\pm$0.30 & -              & 1\\
         73P-C    & 1.02$\pm$0.03 & 3.00$\pm$0.20 & -              & 1\\
         103P     & 1.09$\pm$0.03 & 2.76$\pm$0.08 & 2.12$\pm$0.59 & 1,2,3\\
         C1995O1  & 1.17$\pm$0.13 & 2.45$\pm$0.10 & -              & 1\\
         C1999S4  & 1.16$\pm$0.05 & 2.80$\pm$0.30 & -              & 1\\
         C2000WM1 & 1.12$\pm$0.02 & 2.60$\pm$0.20 & -              & 1\\
         C2001A2  & 1.24$\pm$0.06 & 1.97$\pm$0.20 & -              & 1\\
         C2001Q4  & 1.11$\pm$0.02 & 2.60$\pm$0.30 & -              & 1\\
         C2003K4  & 1.16$\pm$0.04 & 2.47$\pm$0.27 & -              & 1\\
         C2012S1  & 1.14$\pm$0.02 & 2.20$\pm$0.10 & -              & 1\\
         C2013R1  & 1.13$\pm$0.02 & 3.01$\pm$0.49 & -              & 1\\
         C2014Q2  & 1.14$\pm$0.04 & 2.70$\pm$0.76 & -              & 1\\
         \multicolumn{5}{l}{\it Protoplanetary disks} \\
         TW Hya   & -              & $0.2-3.0$    & -         & 4\\
         HD~163296 & -             & -            & 1.8-2.8 & This work\\
         \multicolumn{5}{l}{\it Low-mass protostars} \\
         N1333I4A  & -            & -             & 1.48 & 5\\
         IRAS16293 & -            & -             & 2.50 & 5\\
         L1157     & 1.5$\pm$0.2  & -             & 2.56 & 5,6\\
         L1448C    & -            & -             & 1.31 & 5\\
         N1333I2   & -            & -             & 1.10 & 5\\
         N1333I4B  & -            & -             & 0.21 & 5\\
         VLA1623   & -            & -             & 0.59 & 5\\
         L1527     & -            & -             & 1.00 & 5\\
         \multicolumn{5}{l}{\it High-mass star forming regions} \\
         NGC 6334 I env  & -             & 1.6$\pm$1.0 & -             & 7\\ 
         AFGL 2591 env & -             & 1.9$\pm$0.4 & -             & 8\\
         Orion-KL outf & -             & 3.0 & -             & 9\\
         \multicolumn{5}{l}{\it Photo-dissociation regions} \\
         Orion-Bar & -             & 0.3$\pm$0.2   & 3.$\pm$2   & 10,11\\
         Horsehead PDR & -             & -             & 2.$\pm$0.5 & 12\\
         Horsehead Core & -             & -             & 3.$\pm$0.5 & 12\\
         \multicolumn{5}{l}{\it Pre-stellar clouds} \\
         L1689B    & -             & -             & 1.7 & 5\\
         L1544     & -             & -             & 1.3 & 5\\
     \bottomrule
       \end{tabular}
     \end{center}
       1) \citet{shinnaka2016}, 2) \citet{bonev2013}, 3) \citet{gicquel2014}, 4) \citet{salinas2016}, 5) \citet{jorgensen2005}, 6) \citet{umemoto1999}, 7) \citet{emprechtinger2010}, 8) \citet{choi2015}, 9) \citet{melnick2010}, 10) \citet{choi2014}, 11) \citet{cuadrado2017}, 12) \citet{guzman2011}.
   \end{table}
 }      

\newcommand{\FigALMAobs}{%
\begin{figure*}[t!]
  \centering
  \includegraphics[width=\textwidth]{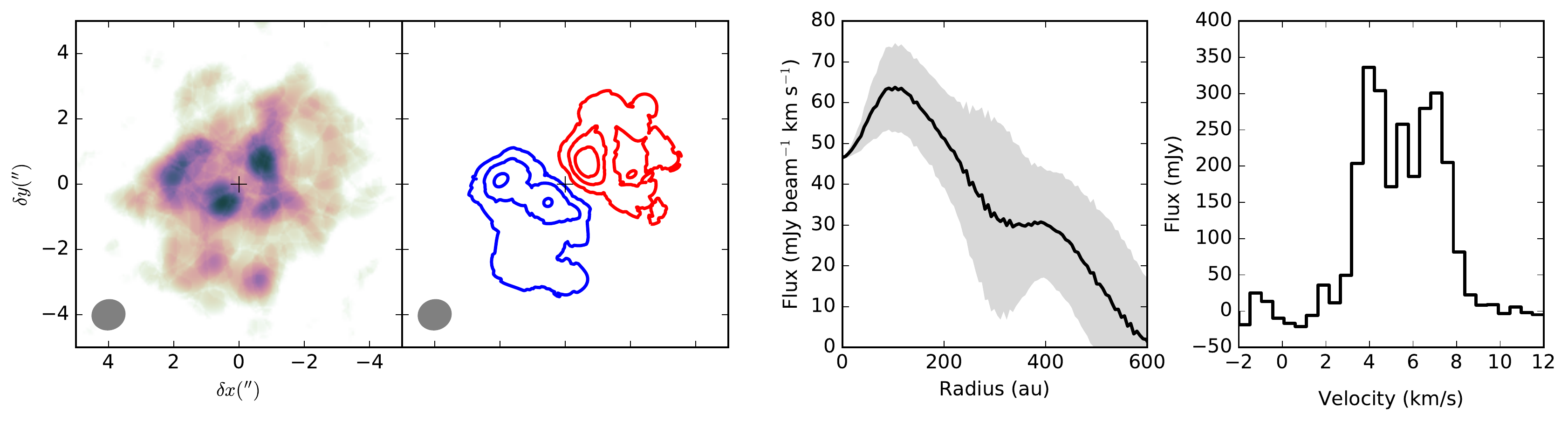}
  \caption{ALMA observations of the \phhco{} $4_{04}-3_{03}$ line. The
    two left panels show velocity integrated maps over the full line
    and over the red- and blue- shifted parts of the line. The beam is
    shown in the bottom left corner. Contours correspond to
    3,7,10,20$\times$rms, where the rms is given in the last column of
    Table~\ref{tab:obs}. The third panel shows the deprojected radial
    profile of the emission computed by azimuthally averaging the
    emission. The gray shaded region shows the standard deviation at
    each radius. The last panel shows the spectra integrated over the
    disk.}
  \label{fig:ALMAobs}
\end{figure*}
}

\newcommand{\FigSMAobs}{%
\begin{figure*}[t!]
  \centering
  \includegraphics[width=\textwidth]{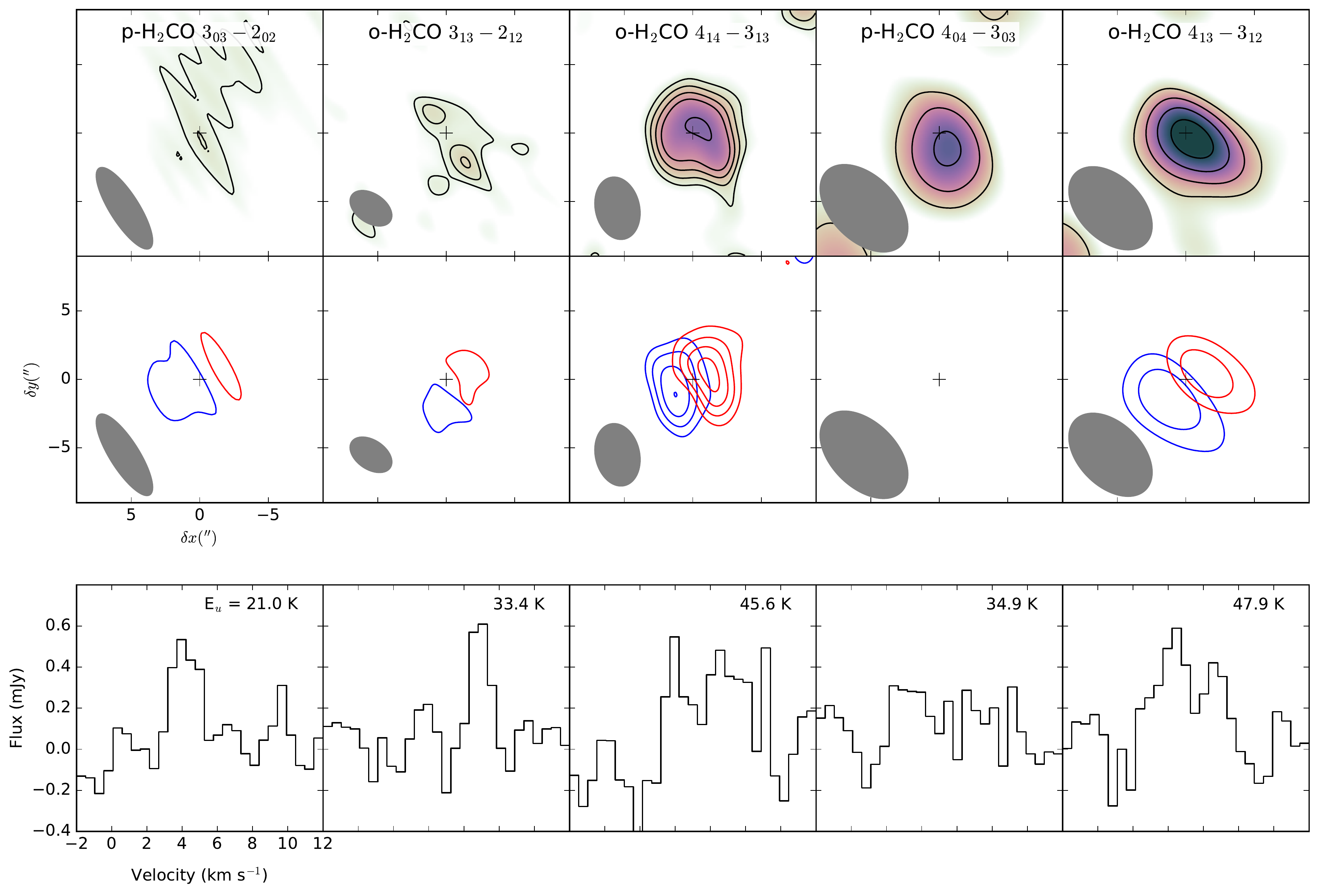}
  \caption{SMA observations of the ortho and para \hhco{} lines. The
    top row show the velocity integrated line emission. The second row
    shows the red- and blue-shifted parts of the emission. Contour
    levels correspond to 3,7,10,20$\times$rms, where the rms is given
    in Table~\ref{tab:obs}. The disk-integrated spectra is shown in
    the bottom row.}
  \label{fig:SMAobs}
\end{figure*}
}

\newcommand{\FigDiskStructModel}{%
\begin{figure}[b!]
  \centering
  \includegraphics[width=0.49\textwidth]{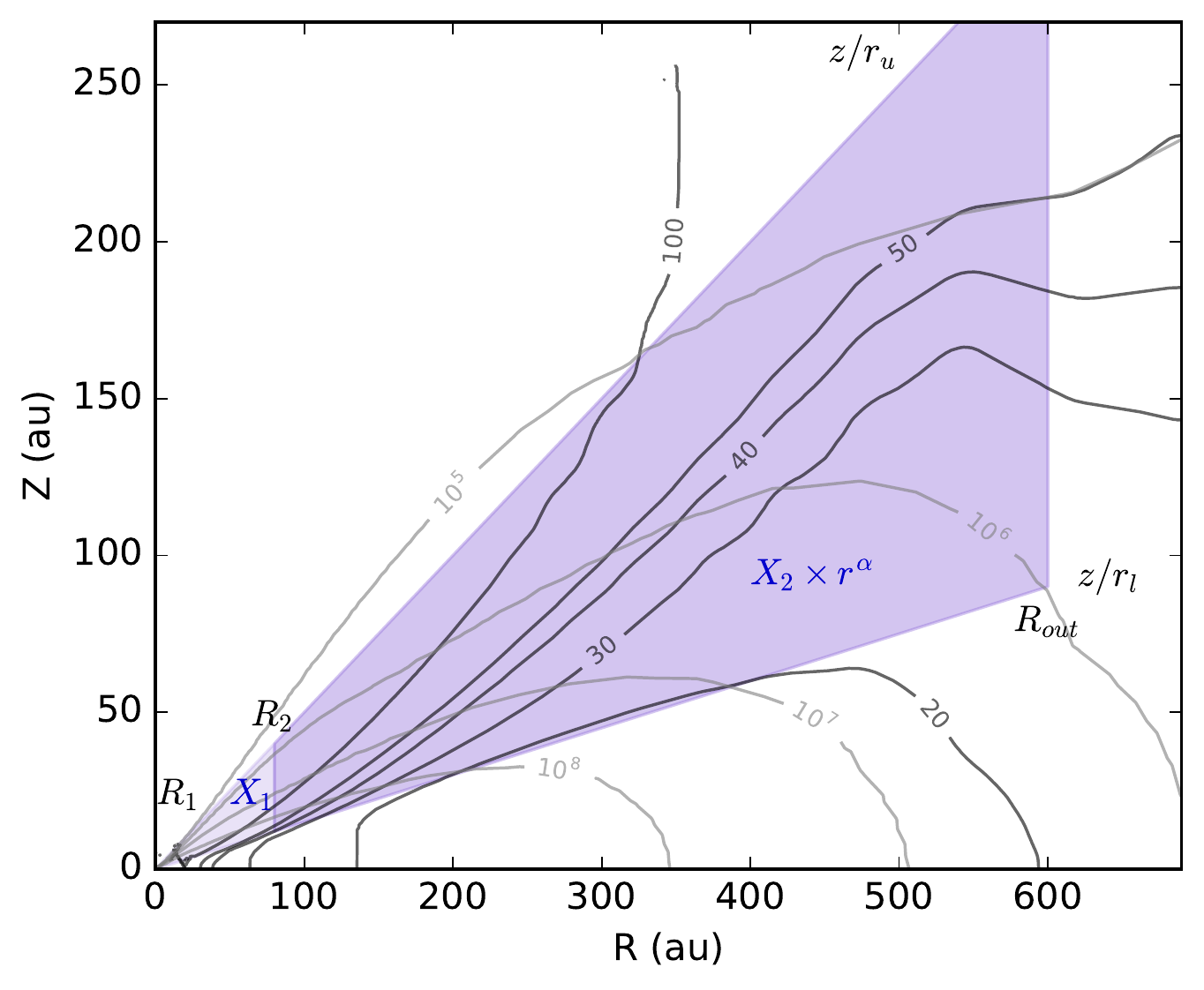}
  \caption{Parameterization of the \hhco{} abundance. The gas
    temperature of the disk (in Kelvin) is shown in black contours,
    and the gas density (in units of $\pccm$) is shown in gray
    contours. The two \hhco{} components considered in our model are
    shown in purple shades. The parameters that are kept constant in
    the model are shown in black, and those that are left free are
    shown in blue.  }
  \label{fig:struc}
\end{figure}
}

\newcommand{\FigRotDiag}{%
\begin{figure}[t!]
  \centering
  \includegraphics[width=0.45\textwidth]{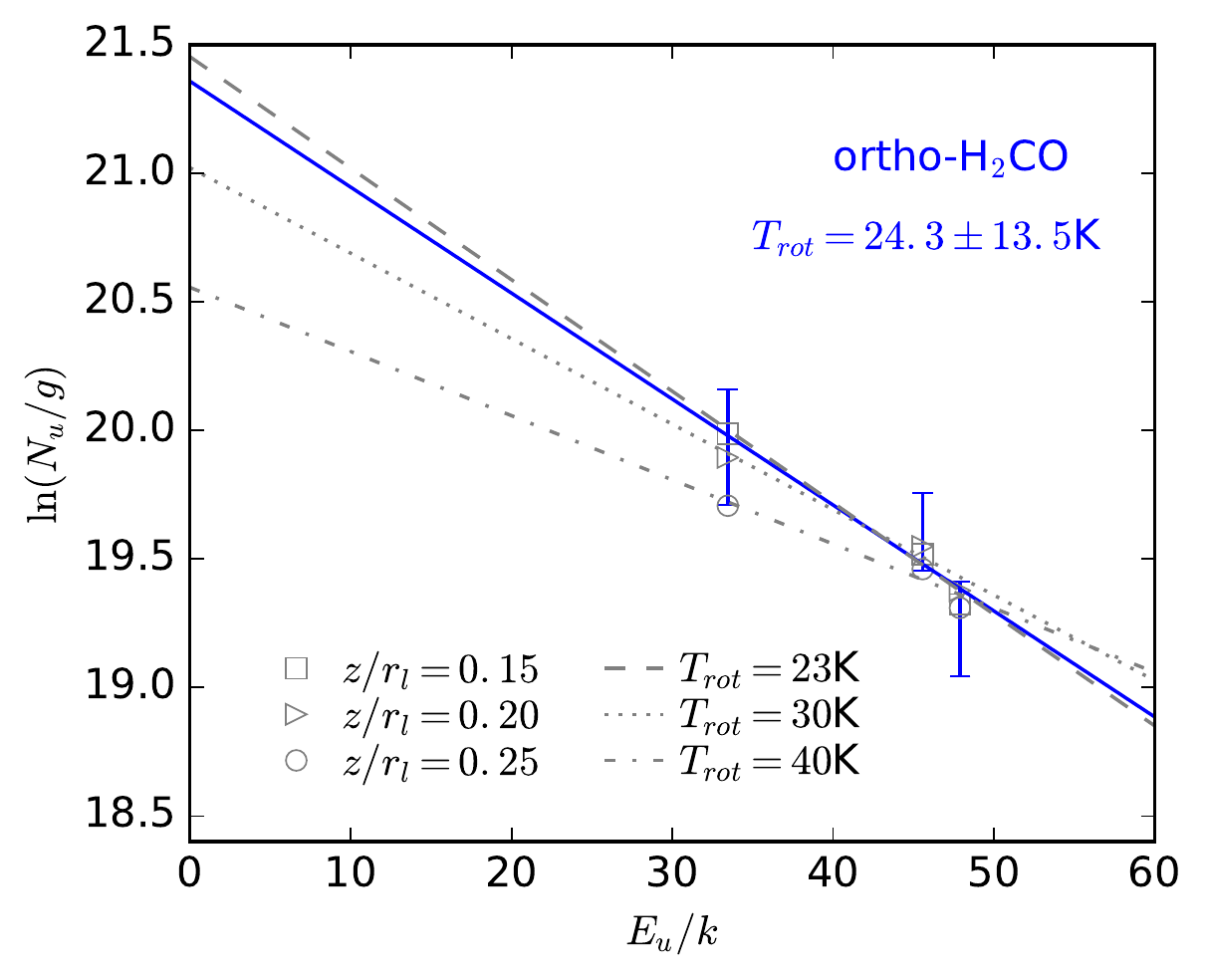}
  \caption{Rotational diagram for \ohhco{}. The blue error bars and solid line correspond to the observations and the gray markers and lines show the best-fit models for the three different lower boundaries considered (see section~\ref{sec:model}).}
  \label{fig:rot-diag}
\end{figure}
}

\newcommand{\FigOPRcomparison}{%
\begin{figure*}[t!]
  \begin{center}
  \includegraphics[width=\textwidth]{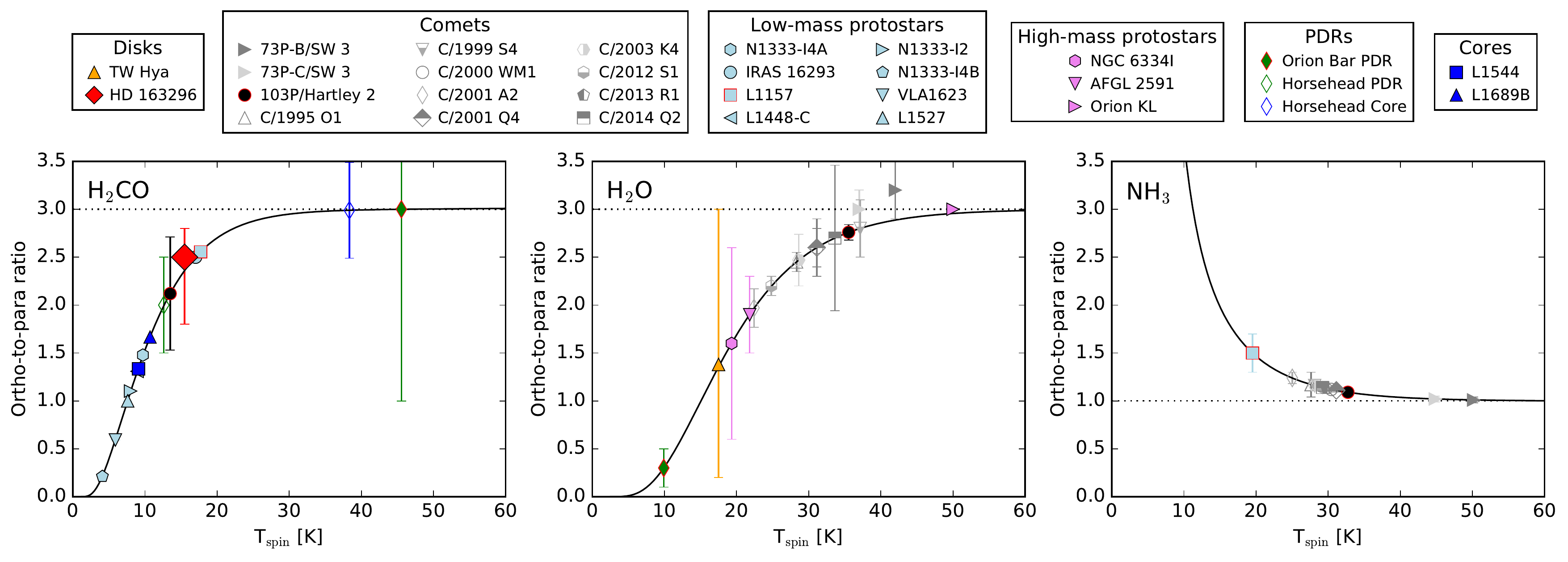}
  \caption{Thermal o/p ratios for \hhco{}, \hho{} and \nhhh{}, as
    given by the Boltzmann distribution of the ortho and para
    levels. The different markers show measurements towards disks$^4$
    (orange and red), comets$^{1,2,3}$ (gray), low-mass
    protostars$^{5,6}$ (light blue), high-mass protostars$^{7,8,9}$
    (pink), PDRs$^{10,11,12}$ (green and blue diamonds) and prestellar
    cores$^{5}$ (blue). Sources with o/p ratios measured for \hhco{}
    and another molecule are highlighted with red edges.
      References: $^1$\citet{shinnaka2016}, $^2$\citet{bonev2013},
      $^3$\citet{gicquel2014}, $^4$\citet{salinas2016},
      $^5$\citet{jorgensen2005}, $^6$\citet{umemoto1999},
      $^7$\citet{emprechtinger2010}, $^8$\citet{choi2015},
      $^9$\citet{melnick2010}, $^{10}$\citet{choi2014},
      $^{11}$\citet{cuadrado2017}, $^{12}$\citet{guzman2011}.}
  \label{fig:opr}
  \end{center}
  
\end{figure*}
}

\newcommand{\FigTriangles}{%
\begin{figure}[t!]
  \centering
  \includegraphics[width=0.15\textwidth]{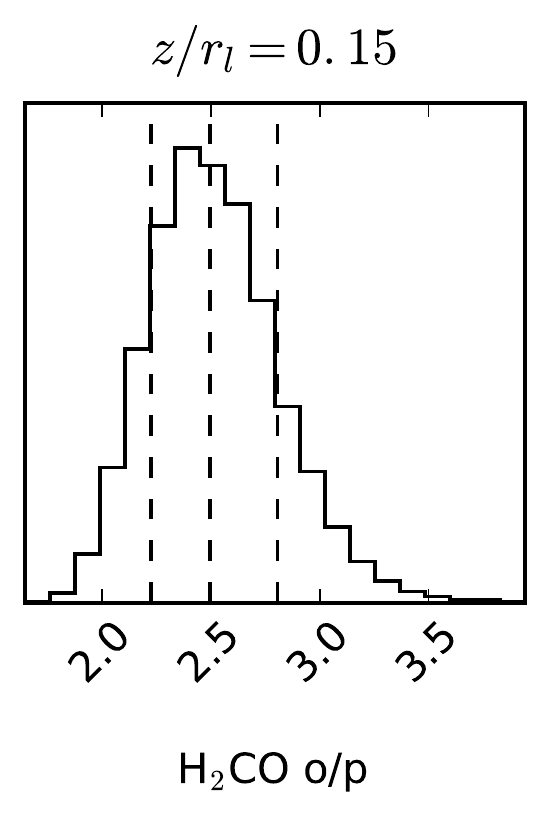}
  \includegraphics[width=0.15\textwidth]{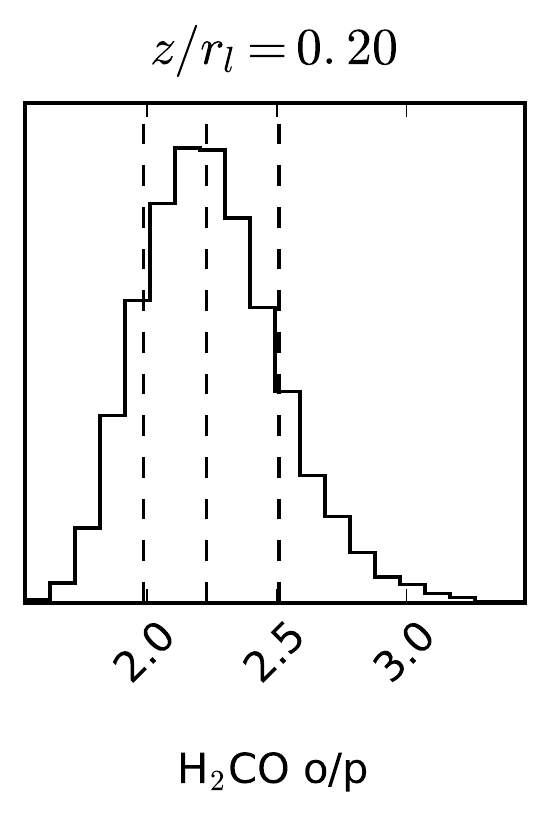}
  \includegraphics[width=0.15\textwidth]{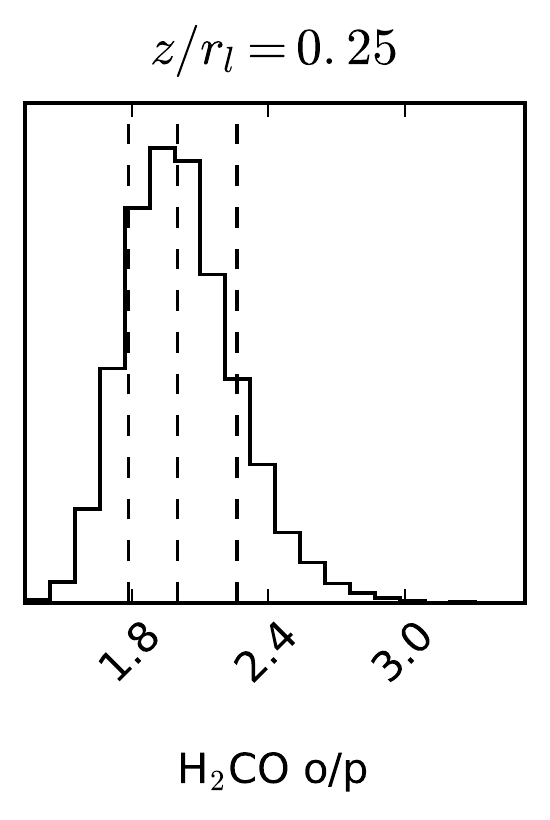}
  \caption{Marginal distribution of the o/p
    ratio in the line modeling for the three lower boundaries
    chosen. The vertical dashed lines mark the median and 1$\sigma$
    uncertainty.}
  \label{fig:triangle}
\end{figure}
}

\newcommand{\FigALMAfit}{%
\begin{figure*}[t!]
  \centering
  \includegraphics[width=\textwidth]{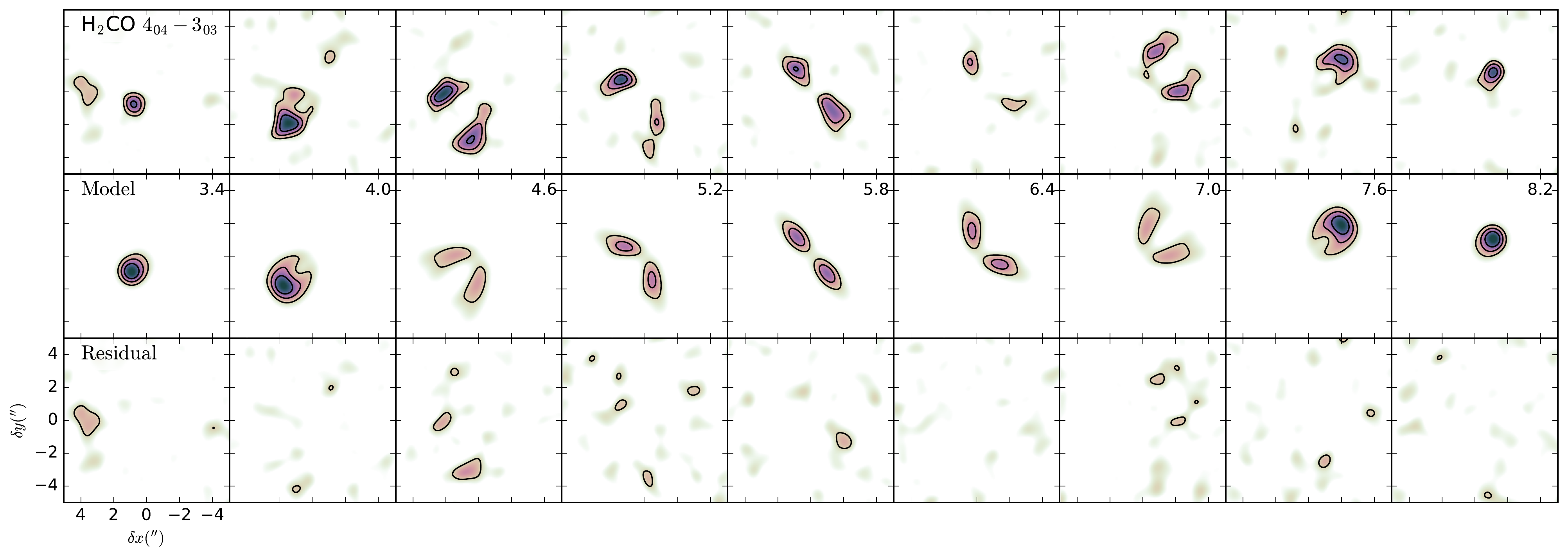}
  \caption{Channel maps of the ALMA \phhco{} line at 290~GHz. The
    observed line is shown in the top row, while the best fit model is
    shown in the middle row, and the residuals are shown in the bottom
    row. The velocity in \kms{} of each panel is shown at the top right in the middle row. Contours are 3, 5 and 7$\sigma$.}
  \label{fig:res}
\end{figure*}
}


\section{Introduction}

\TabObs{}
\FigALMAobs{}

Formaldehyde (\hhco) is a simple organic molecule that is easily
detected in the interstellar medium (ISM). Being a slightly asymmetric
rotor, it is an excellent tracer of the gas density and
temperature. In addition, \hhco{} could potentially be used to trace
snowlines in protoplanetary disks, an important parameter for gas
giant planet formation \citep{qi2013}. However, to correctly interpret
the observations it is necessary to understand the dominant formation
pathway of \hhco{}. Understanding its formation in disks is of
especial importance because disks provide the material for new
planetary systems. \hhco{} can form both in the gas-phase and on the
surface of dust grains \citep[e.g.,][]{fuchs2009}, but the relative
importance of these two mechanisms is poorly constrained in many
interstellar environments, including disks. A possible way to
discriminate between the two is to measure the ortho-to-para (o/p)
ratio.

Molecules that contain two or more hydrogen atoms, such as H$_2$,
H$_2$O, \hhco{} and NH$_3$, exist in two isomeric forms: ortho
(hydrogen nuclei spin are parallel) and para (hydrogen nuclei spin are
anti-parallel). In thermal equilibrium, the relative population
between the two isomers is given by:
\begin{equation}
\rm{o/p} = \frac{\sum g^{ortho}_i \exp(-\frac{E^{ortho}_i}{k_B\Tspin})}{\sum g^{para}_i \exp(-\frac{E^{para}_i}{k_B\Tspin})},
\end{equation}
where $g_i$ and $E_i$ are the level degeneracies and energies,
respectively, $k_B$ is the Boltzmann constant and \Tspin{} is the spin
temperature.

If the formation temperature is high compared to the energy difference
between the two species, the molecule will form with an o/p ratio
equal to the statistical value, which corresponds to the statistical
weight ($g_i$) ratio between the ground states of the ortho and para
species. At thermal equilibrium, low-temperature formation pathways
should deviate from the statistical value, and result in a lower
$\Tspin$. In the case of \hhco{} (and \hho{}), when it forms in the
warm gas-phase, an o/p ratio of 3 is expected. In contrast, a lower
o/p ratio could result if \hhco{} forms on the surface of dust grains
\citep{kahane1984,dickens1999}.

A broad range of o/p ratios have been measured in comets and in the
ISM, for \hhco{}
\citep[\eg{},][]{kahane1984,maret2004,jorgensen2005,guzman2011} and
other species, like H$_2$Cl$^+$ \citep{legal2017} and NH$_3$
\citep{persson2012}. Observations of o/p ratios for water and ammonia
have been commonly used to deduce a cold origin of the cometary
volatiles in the Solar Nebula \citep{mumma2011}. However, the
interpretation of o/p ratios is now disputed, as recent laboratory
experiments of ices are not able to reproduce the low \hho{} o/p
ratios that are observed in comets and in the ISM \citep{hama2016}.

Measurements of o/p ratios in Solar nebula analogs, \ie{}
protoplanetary disks, could help to elucidate the meaning of o/p
ratios in solar system bodies. So far, only the water o/p ratio has
been measured in TW Hya using Herschel observations, where a large
range of possible values was found
\citep[0.2-3.0;][]{hogerheijde2011,salinas2016} depending on the disk
model. \hhco{} is particularly well suited for such studies because
its lines are bright in disks at millimeter wavelengths, and several
lines of both its ortho and para states can now be easily detected
with interferometers. Indeed, \hhco{} has been detected in several
disks
\citep{aikawa2002,oberg2010,oberg2011,qi2013,loomis2015,oberg2017,carney2017},
but so far too few lines have been observed to put any constraints on
the o/p ratio. In this paper, we provide the first measurement of the
disk-averaged \hhco{} o/p ratio in a protoplanetary disk.

Our target, the HD~163296 disk, is inclined by $48.5\deg$ (as measured
by CO and dust continuum emission), and the position angle (defined
from North to West) is 132$\deg$ \citep{isella2007}. The central star
has a mass of 2.3~\Msun{}, a disk mass of $0.01$~\Msun{}, and it is
located $101.5\pm1.2$~pc away \citep{gaia2018}. For consistency with
previous analysis, we use the old distance of 122 pc
\citep{vandenAncker1998} throughout the paper. The size of the dust
disk at milimeter wavelengths is $\sim250$~au and the CO emission
extends to $\sim550$~au \citep{isella2016}. The location of the CO
snowline has been measured with ALMA observations of N$_2$H$^+$
\citep{qi2015}, providing an additional constraint on the possible
\hhco{} origin in this particular disk. HD~163296 is thus particularly
well suited as a test case for this work.

The observations and data reduction are presented in
section~\ref{sec:obs}. In section~\ref{sec:results} we describe the
\hhco{} spatial distribution seen by the ALMA observations, and derive
a rotational temperature from the SMA observations. In
section~\ref{sec:model} we describe the model used to reproduce the
line emission, and extract the disk-averaged \hhco{} o/p ratio. A
discussion is presented in section~\ref{sec:discussion}, and a we
summarize the results in section~\ref{sec:conclusions}.

\FigSMAobs{}

\section{Observations and data reduction}
\label{sec:obs}

We obtained Atacama Large Millimeter/submillimeter Array (ALMA)
observations of the $J=4-3$ para \hhco{} line, and Submillimeter-Array
(SMA) observations of the $J=3-2$ and $J=4-3$ lines of both ortho and
para \hhco{} towards the disk around the Herbig Ae star HD~163296. A
description of the observations and data reduction is given next.

\subsection{ALMA}

We observed the \hhco{} $4_{04}-3_{03}$ line at 290.623 GHz with ALMA
in May 2014 during Cycle 1 (project ADS/JAO.ALMA\#2012.1.00681; PI:
C. Qi.) with 32 antennas. The Band 7 observations have baselines
lengths spanning between 27 and 650~m, and a total on-source time of
45.5~min. The correlator was configured with four spectral windows,
covering the N$_2$H$^+$ $J=3-2$, DCO$^+$ $J=4-3$ and the \hhco{}
$4_{04}-3_{03}$ lines, as well as a window dedicated to continuum. A
detailed description of the observations and the results of the
N$_2$H$^+$ $J=3-2$ and DCO$^+$ $J=4-3$ lines are given in
\citet{qi2015}. The \hhco{} line was covered with a spectral window of
58~MHz bandwidth and 15~kHz channel width. To calibrate the amplitude
and phase temporal variations, the quasar J1733-1304 was
observed. 

We used the standard calibration performed by the ALMA staff, and
further self-calibrated the data in phase and amplitude to improve the
signal-to-noise ratio. The continuum subtracted visibilities were
cleaned in \CASA{} 4.7, using the \CLEAN{} algorithm with Briggs
weighting of 2. The data were regridded to a spectral resolution of
0.6\kms{}. We applied a Gaussian taper of $1''$ to improve the
signal-to-noise. In addition, we created a mask to help the cleaning
process, by selecting regions in each channel with line emission that
is consistent with the expected Keplerian rotation of the disk. The
resulting noise properties and beam information are summarized in
Table~\ref{tab:obs}.

\subsection{SMA}

We observed five \hhco{} lines with the SMA. The \phhco{} lines at
218~GHz and 290~GHz, and the \ohhco{} line at 300 GHz were observed in
March 2014 with 7 antennas in the sub-compact (SUB) array
configuration. The 218~GHz line was observed in one spectral setting,
with a spectral window of 104~MHz and 406~kHz channel width. The 290
and 300~GHz lines were covered in another setting, with two spectral
windows of 104~MHz and 406~kHz channel width each. The quasars NRAO~530
and $1744-312$ were observed to calibrate amplitude and phase variations,
and the frequency bandpass was calibrated using observations of
3C279. Titan and Mars were used as flux calibrators. We also include
observations of the \ohhco{} $4_{14}-3_{13}$ and $3_{12}-2_{11}$ lines
at 281.527 and 225.698~GHz, respectively, which have been presented
in \citet{qi2013}. A detailed description of the observations can be
found in \citet{qi2013}. For consistency with the other \hhco{} lines,
we present an independent imaging in this paper.

The SMA data were first calibrated with the \MIR{} software package
using standard
procedures\footnote{\url{https://www.cfa.harvard.edu/~cqi/mircook.html}}. The
calibrated data were then exported to \CASA{} and cleaned in the same
way as the ALMA line. The data were regridded to a common spectral
resolution of 0.6\kms{}. A robust parameter of 2 was used to optimize
the sensitivity. To help the cleaning process, we created circular
masks for each line, that was the same for each channel. The
radius of the mask was chosen to cover the line emission in all
channels, and varied for the different \hhco{} lines. The resulting
rms noise levels and observational parameters are summarized in
Table~\ref{tab:obs}.

\section{Observational results}
\label{sec:results}

\subsection{\hhco{} spatial distribution}

Fig.~\ref{fig:ALMAobs} shows the ALMA observations of the \phhco{}
$4_{03}-3_{03}$ line. The two left panels show the velocity integrated
map, integrated over the full velocity range (left) and over the blue-
and red-shifted part of the line (right). The third panel in
Fig.~\ref{fig:ALMAobs} shows the azimuthally-averaged deprojected
profile of the \phhco{} $4_{03}-3_{03}$ line emission. It was obtained
by extracting elliptical anulii in the unclipped moment-zero map,
taking into account the inclination and position angle of the
disk. The right-most panel shows the disk-integrated spectra, which
was obtained using an elliptical mask. The integrated spectra shows a
double-peaked profile, typical of Keplerian rotation of an inclined
disk. The emission presents a ring-like spatial distribution, with a
depletion of \hhco{} emission at the disk center, a peak of emission
at intermediate radii ($\sim100$~au), and either a second but weaker
peak or a plateau in the outer disk near $\sim400$~au. This second
bump is located farther out in the disk than the milimeter dust edge
near $\sim250$~au \citep{isella2016}. \citet{carney2017} presented
higher-angular resolution observations (0.5'') of the lower energy
($E_u=21$~K) $3_{03}-2_{02}$ line, which has a similar radial profile.

The observed \hhco{} distribution is not unique to the HD~163296
disk. Spatially resolved observations of the low-energy
($E_u\sim30$~K) \hhco{} $3-2$ line in TW~Hya also show a depression of
the emission at the disk center and a bump in the outer disk
\citep{oberg2017}. In contrast, observations of the higher-energy
($E_u\sim60$~K) $5-4$ line in DM~Tau and TW~Hya present centrally
peaked emission profiles \citep{loomis2015,oberg2017}. Such different
morphologies of the $3-2$ and $5-4$ lines are well explained by the
presence of hot \hhco{} close to the star followed by a lack of
\hhco{} until cold \hhco{} formation becomes possible in the outer
disk. In summary, the observed \hhco{} morphology toward HD~163296 is
consistent with previous \hhco{} disk observations, suggesting that it
is a good object for a case study on \hhco{} chemistry in disks.


\subsection{\hhco{} excitation temperature}
\label{sec:trot}

The SMA observations are shown in Fig.~\ref{fig:SMAobs}. The top row
shows the velocity integrated maps of the five \hhco{} lines, while
the red- and blue-shifted parts of the emission are shown in the
second row. The bottom row shows the disk-integrated spectra, obtained
using a circular mask with a radius of 6''. Although the emission
looks washed-out in the moment-zero maps for some of the lines, visual
inspection of the individual channels show line emission consistent
with the expected Keplerian rotation of the disk. The two higher
energy \hhco{} lines ($E_u>45$~K) are centrally peaked. The peak
emission of the remaining lines with $E_u<40$~K are not centered,
suggesting ring-like emission patterns or a drop of the emission in
the inner disk consistent with the higher-angular resolution ALMA
observations. 

We extracted disk-integrated fluxes using a circular mask with a
radius of 6''. The fluxes are reported in Table~\ref{tab:fluxes}. The
uncertainty in the flux was estimated by taking the standard deviation
of 500 simulated flux measurements from regions free of signal, using
the same circular mask but centered at random positions in the
map. The fluxes of the 290~GHz line observed with ALMA and SMA are
consistent within the uncertainties. We use the disk-integrated fluxes
to estimate an excitation temperature for \hhco{}.

The critical densities of the observed \hhco{} lines, defined as the
density at which the spontaneous radiative de-excitation rate
($A_{ul}$) is equal to the collisional de-excitation rate
($\gamma_{ul}$), are given in Table~\ref{tab:fluxes} for a temperature
of 20~K. For a higher temperature of 60~K, the critical densities
increase by less than 20\%. When the gas density is comparable or
larger than the critical density, the emission is thermalized and the
excitation temperature is equal to the gas kinetic temperature. When
the gas density is lower, then the emission becomes sub-thermally
excited and $\Tex < \Tkin$. Given the high ($>10^6 \pccm$) gas
densities in protoplanetary disks, in particular closer to the
midplane (see Fig.~\ref{fig:struc}), we expect the bulk of the
emission to be thermalized.

Using the three ortho lines we constructed a rotational diagram
(Fig.~\ref{fig:rot-diag}). We obtain a rotational temperature of
$24\pm14$~K. This temperature is consistent with the temperature of
24~K derived by \citet{carney2017}, from the \phhco{} $3_{03}-2_{02}$,
$3_{22}-2_{21}$ and $3_{21}-2_{20}$ lines, using a match-filter
technique to extract the fluxes of the two weaker lines.  The derived
rotational temperature also matches the CO freeze-out temperature
found by \cite{qi2015} for this disk, using N$_2$H$^+$ and C$^{18}$O
observations. This suggest that the \hhco{} emission at the large
scales, to which the SMA is sensitive, arises predominantly from the
cold layers close to the midplane, where CO starts to freeze-out onto
dust grains and can be hydrogenated to form \hhco{}. In the next
section, we use this measured rotational temperature to inform our
model on the distribution of \hhco{} in the disk.

\TabFluxes{}
\FigRotDiag{}



\FigDiskStructModel{} 
\TabModelParam{}
\FigALMAfit{}

\section{Line modeling}   
\label{sec:model}

In this section we aim at measuring the disk averaged \hhco{} o/p
ratio in the HD~163296 disk. For this, we use the disk physical
structure of \citet{qi2011}, that was constrained by fitting the
spectral energy distribution, the millimeter dust continuum, and
multiple CO and CO isotopologue lines. The adopted gas density and
temperature structure is shown in Fig.~\ref{fig:struc} with gray and
black contours, respectively. We first model the abundance structure
of \hhco{} using the high angular resolution observations of the
\phhco{} line. Keeping the spatial distribution of \hhco{} constant,
we then fit the five \hhco{} lines observed with the SMA to find the
best-fit o/p ratio.

\subsection{\hhco{} abundance structure}

Based on chemical model predictions of disks, the distribution of
\hhco{} can be separated in two main components: 1) A warm component
towards the center of the disk, that is mainly produced by gas-phase
chemistry \citep[\eg{},][]{loomis2015}, and 2) a colder component
located in the layers above the midplane in the outer disk, where the
\hhco{} formation is dominated by grain surface chemistry
\citep[\eg{},][]{oberg2017}. In the midplane, \hhco{} remains frozen
on the surfaces of dust grains. Following this, we parameterize the
abundance structure of \hhco{} assuming the two components displayed
in Fig.~\ref{fig:struc}. The first component is located in the inner
disk, between $R_1$ and $R_2$, with constant abundance $X_1$. The
second component is located between $R_2$ and $R_{out}$, and has an
abundance described by a power-law,
$X=X_2(r/100~\mathrm{au})^\alpha$. Both components have lower and
upper boundaries, $(z/r)_l$ and $(z/r)_u$.

Given the uncertainties when modeling only one transition, we fixed
some of the parameters. The fixed parameters are $R_1, R_2, R_{out},
(z/r)_{l}$ and $(z/r)_u$; and only $X_1, X_2,$ and $\alpha$ are left
free in the line fitting. The adopted values in the model are given in
the left column of Table~\ref{tab:param}. The values for $R_1$ and
$R_2$ are chosen based on predictions from chemical models of this
disk (Cleeves et al., private communication). $R_{out}$ is fixed to
600~au, corresponding to the extent of the \hhco{} emission. We note
that due to the low-angular resolution of the SMA observations, the
inferred disk-averaged o/p ratio (see section~\ref{sec:fit-op}) is not
very sensitive to variations in the position of the inner component
(\ie{}, $R_1$ and $R_2$). We consider three different values (0.15,
0.20 and 0.25) for the lower boundary, $(z/r)_l$, to investigate the
impact of the adopted height of the \hhco{} disk emission layer on the
derived o/p ratio. The upper boundary, $(z/r)_u$, is fixed to 0.5; we
tried different values, and found that the results were not sensitive
to this paramter.

We use a Bayesian approach to find the best fit of the model to the
ALMA visibilities. For this, we first create a synthetic observation
of the \phhco{} line emission using the \RADMC{} package
\citep{dullemond2012} to compute the level populations, assuming the
gas is under local thermal equilibrium (LTE). We then compute the
Fourier Transform of the modeled line emission, using the \vissample{}
Python package\footnote{\url{https://pypi.python.org/pypi/vis sample}} that
re-projects the modeled visibilities on the observed $u-v$ points. We
then compute the weighted difference between model and observations,
for the real and imaginary parts of the complex visibility, which
gives the likelihood of the model. This process is repeated by
sampling the posterior distribution with the MCMC method implemented
in the \emcee{} package by \citet{Foreman-Mackey2013}.

The best fit values of the model parameters are listed in the right
columns in Table~\ref{tab:param}, for the three different lower
boundaries we consider. Because we only model one \phhco{} line, the
abundance we derive corresponds to that of \phhco{}. The abundances
listed in Table~\ref{tab:param}, however, correspond to the total
\hhco{} abundance, including both ortho and para, and are computed
assuming a uniform o/p ratio of 3. We allow this parameter to vary in
the next section. The best-fit is found for the model with
$(z/r)_l=0.15$. Fig.~\ref{fig:res} shows channel maps of the \hhco{}
line observed with ALMA (top row), together with the best fit model
(middle row), and the residuals (bottom row). The inferred abundances
in the inner $\sim200$~au change considerably (by more than one order
of magnitude) depending on the adopted value for the lower
boundary. The \hhco{} abundance in the outer disk, given by the second
component, change by a factor of $\sim3-7$ (at $300-500$~au) within
the three sets of models.

Motivated by the bump observed in the radial profile of the \hhco{}
ALMA line (see Fig.~\ref{fig:ALMAobs}), we also explored the presence of a
third component in the outer disk that is closer to the midplane. For
this, we ran another set of models including a third component of
constant abundance between 350 and 600~au, and upper boundary equal to
the lower boundary of the second component (\ie{}, $0<z/r<(z/r)_l$). The
best-fit abundance of this third component remained low compared to
the two other components no matter the adopted $(z/r)_l$ value,
suggesting that the contribution of such a third cold component to the
overall \hhco{} emission is small. We note that these models improved
slightly the fit of the ALMA \hhco{} line but resulted in similar
results for the SMA observations. 

\subsection{Disk-average \hhco{} o/p ratio}
\label{sec:fit-op}

\FigTriangles{}

We model the line emission of the two para and three ortho lines of
\hhco{} in the same way we did for the ALMA observations. Given the
lower angular resolution of the SMA observations, we keep the \hhco{}
abundance structure fixed, and vary only the o/p ratio. We assume a
uniform o/p ratio across the disk. In order to take into account for
differences in the absolute fluxes between the ALMA and SMA data due
to absolute flux calibration uncertainites of 10-20\%, we introduce a
{\it scale} factor for the total \hhco{} abundance in the
disk. That is, the total \hhco{} abundance derived in the previous
section is multiplied by this {\it scale} factor, which is a free
parameter in the SMA line fitting. 

The resulting best-fit parameters are shown in the bottom rows of
Table~\ref{tab:param}. Fig.~\ref{fig:triangle} shows the posterior
probability distribution for the \hhco{} disk-averaged o/p ratio.  The
{\it scale} factor is $1.32-1.76$ (depending on $(z/r)_l$), and
increases somewhat with increasing lower boundary. These values are of
the expected magnitude when considering the differences in measured
fluxes with ALMA and SMA (see section~\ref{sec:trot}).

\FigOPRcomparison{}

The inferred o/p ratio changes depending on the assumed lower boundary
but remains consistently lower than 3, the high-temperature value. The
highest o/p value (2.5) is obtained when \hhco{} is located closer to
the midplane, and lower values (2.2 and 2.0) are obtained when \hhco{}
is closer to the disk surface. The possible values for the \hhco{} o/p
ratio are $1.8-2.8$ (within 1$\sigma$), corresponding to spin
temperatures of $11-22$~K. Fig.~\ref{fig:rot-diag} shows the
rotational diagram of \ohhco{}, and compares the observed and best-fit
model integrated fluxes for the three $(z/r)_l$ choices. The best-fit
is obtained for $(z/r)_l=0.15$, which gives an o/p ratio of
$2.50^{+0.31}_{-0.27}$, corresponding to $\Tspin=17^{+6}_{-3}$~K. This
is the only model resulting in consistent spin and rotational
temperatures. However, we cannot rule out any of the vertical
distribution choices based on the SMA data.

We note that this corresponds to a disk-averaged o/p ratio. Most
likely, this value changes across the disk, both radially and
vertically. Our inferred o/p value thus suggest that a considerable
amount of the gas in the HD~63296 disk has an \hhco{} o/p ratio that
is lower than 3. Higher-angular resolution observations are needed to
disentangle regions in the disk with low o/p ratios from regions
presenting a high-temperature statistical value.

\section{Discussion}
\label{sec:discussion}

It is believed that o/p measurements provide information on the
formation conditions of the molecule. In this section we aim at
comparing the observed \hhco{} o/p ratio in the HD~163296
protoplanetary disk with values observed in the ISM and solar system
bodies (for different species) to better understand how and if o/p
ratios can be used to elucidate the formation pathway of \hhco{}.  We
also discuss the meaning of the o/p value in light of the new
observations and what additional observations can resolve existing o/p
puzzles.

\subsection{Observed o/p ratios in the ISM and comets}

The \hhco{} o/p ratio has been measured in a variety of interstellar
environments and one comet. Fig.~\ref{fig:opr} presents a compilation
from the literature of measured \hhco{} o/p ratios. In the subsequent
panels we present the same data on \hho{} and \nhhh{}. We highlight
with red edges those sources where the o/p ratio has been measured for
\hhco{} and another species. The solid curves in Fig.~\ref{fig:opr}
correspond to the predicted o/p curves given by the Boltzmann
distribution of the ortho and para energy levels. The observations are
also listed in Table~\ref{tab:oprs}.

A range of \hhco{} o/p ratios is found in the ISM. Low-temperature
($\Tspin<30$~K) values have been observed in different environments,
such as the envelopes around low-mass stars \citep{jorgensen2005},
outflows \citep{dickens1999}, and the Horsehead PDR
\citep{guzman2011}. \hhco{} o/p ratio consistent with the
high-temperature value of 3 have been observed towards quiescent
cores, such as the cold ($\sim20$~K) UV-shielded core located just
behind the Horsehead PDR \citep{guzman2011}, and the Orion-Bar PDR
\citep{cuadrado2017}. Because the Orion-Bar ($\Td \sim 150$~K) is a
much stronger UV-illuminated PDR, dust grains at the cloud edge are
free from ice mantles, in contrast to the Horsehead PDR where icy
mantles can exist ($\Td \sim 20-30$~K). \cite{cuadrado2017} have thus
suggested that \hhco{} forms predominantly in the gas-phase in the
Orion-Bar, while it forms on ices and is later non-thermally desorbed
into the gas-phase in the Horsehad PDR. However, recent results from
laboratory experiments cast doubt on this interpretation (see section
5.2 for a discussion).

The \hhco{} o/p ratio has been measured in one comet, 103P/Hartley 2,
where a value of $2.1\pm0.6$ was found, corresponding to
$\Tspin\sim13.5^{+6.7}_{-3.3}$~K \citep{gicquel2014}. This o/p value
is consistent with our disk measurement of $1.8-2.8$, but the spin
temperature is lower compared to what has been deduced from a sample
of \nhhh{} and \hho{} o/p ratio comet measurements;
\citet{shinnaka2016} measured the \nhhh{} o/p ratio in 26 comets, and
found values of $1.1-1.2$, corresponding to a spin temperature of
30~K. The \hho{} o/p ratio has been measured towards about half of
them. The similar \hho{} and \nhhh{} $\Tspin$ suggests that they share
a common chemical history and/or they follow a similar thermalization
mechanism, while \hhco{} either forms or thermalizes
differently. Observations of \hhco{} o/p ratios in more comets are
needed to confirm this is a general behavior.

Towards disks, only the water o/p ratio has been measured so far and
only in one source, TW~Hya, but the range of estimated ratios is too
wide to provide any constraints. \cite{hogerheijde2011} first reported
an \hho{} o/p ratio of $0.77\pm0.07$, but later \cite{salinas2016}
explored different disk models and found values between 0.2 and 3.
The disk-averaged \hhco{} o/p ratio measured in the HD~163296 disk is
consistent with the \hho{} o/p ratio measured in TW~Hya. 

\TabOPRs{}

\subsection{The meaning of o/p ratios and future directions}


It has long been thought that the o/p ratio upon formation is
preserved and \Tspin{} should reflect the formation temperature. This
is because the ortho-para inter-conversion by spontaneous radiative
transitions and non reactive collisions are extremely slow
\citep[\eg{}][]{pachucki2008,tudorie2006}, so the o/p ratio upon
formation should be preserved.


In general, the o/p ratio upon formation of any molecule will depend
on its formation mechanism, the spin selection rules, and the o/p of
the reactants. Then, the ortho and para symmetry species can relax to
the lowest energy rotational levels through spontaneous
emission. Ortho-para conversion can occur via reactive collisions, for
example with hydrogen atoms, in which a proton is exchanged. This
process, known as equilibration, will compete with the destruction of
the molecule. Whether equilibration or destruction dominates will
depend on the specific rates, and will determine the resulting o/p
ratio of the molecule. A good description of the different
interstellar processes that determine the o/p ratio is given by
\citet{herbst2015}, for the example of the water cation.

In the specific case of \hhco{}, the dominant gas-phase formation
pathway are reactions of atomic oxygen with CH$_3$. The destruction of
\hhco{} will depend on the conditions of the gas. In the presence of
strong UV fields (i.e., in the disk surface layers and inner disk),
\hhco{} will mainly be photo-dissociated, while in UV-shielded regions
(i.e., in the disk midplane and outer disk) it will mainly be
destroyed by reactions with ions \citep{guzman2011}. The original
\hhco{} o/p ratio will thus depend on the o/p ratio of CH$_3$. CH$_3$
forms from the dissociative recombination of CH$^+_5$ with electrons,
where CH$^+_5$ in turn forms from a series of reactions involving
H$_2$, starting from C$^+$. The fraction of ortho and para formed in
each of these reactions (the branching ratios) will depend on the
details of the formation mechanism, which are not fully understood.
When a di-hydrogenated molecule like \hhco{} forms in the gas-phase
through full proton scrambling mechanisms, the resulting o/p ratio
should reflect the o/p ratio of its parent molecule.  Gas-phase
formation in cold gas that is enriched in p-\hh{} should thus result
in o/p ratio $<3$, while formation in the warmer gas should result in
the high-temperature limit. However, if full proton scrambling
mechanisms are not possible/allowed and only restrictive mechanisms
pertain (such as H-abstraction), then even a cold gas-phase formation
pathway could result in a statistical ratio of 3 resulting in a
complete loss of chemical history information. The exact calculations
require an accurate treatment of the selection rules (see for example
\citet{gerlich2006} for the case of H$_3^+$, and \citet{legal2017} for
the case of H$_2$Cl$^+$). These do not yet exist, and theoretical
studies focused on \hhco{} are needed to determine which of these
processes dominate, and thus whether the thermal history of the gas
can be traced with \hhco{} o/p ratio measurements.

The resulting o/p ratio from grain surface formation is even less
clear, although it is often suggested that the \Tspin{} will reflect
the temperature of the ice \citep[\eg{},][]{kahane1984}.  Only
recently have the nuclear-spin conversion on ices, by thermal and
non-thermal desorption processes, been investigated. Laboratory
experiments have shown that after formation on water ices (at
$<10$~K), \hh{} has the statistical o/p ratio of 3. But the o/p ratio
was found to decrease substantially if \hh{} is re-trapped in the ice
\citep{watanabe2010}. The resulting \hh{} o/p ratio thus seems to
depend on how long the molecule resides in the ice. More recently,
\citet{hama2016} investigated the o/p conversion of water on ices, and
found that the statistical value of 3 is preserved after the water
ices are thermally desorbed into the gas-phase (when heating the ices
to $\sim150$~K), no matter if the \hho{} ice is formed by
vapor-deposition or it is formed in-situ at 10~K. One could argue that the
water o/p ratio will equilibrate to the (high) temperature during
thermal desorption, while it can remain low in the case of
photo-desorption. However, \citet{hama2016} also investigated the case
of photo-desorption, and found that the desorbed water also has an o/p
ratio of 3. This unexpected result raised the question of what is the
real meaning of the low o/p ratios measured in comets, star and planet
forming regions.

Assuming that \hhco{} behaves similarly to water, it is possible that
the \hhco{} o/p ratio is not indicative of the ice formation
temperature, but could instead provide information on the desorption
mechanism and/or the physical conditions of the gas at the time of
desorption. Indeed, the spin temperature inferred from \hhco{} in
HD~163296 agrees well with the expected gas temperature in the regions
where the bulk of \hhco{} is expected to form. More laboratory
experiments and theoretical predictions are needed to better understand
the different formation and desorption processes and how they affect
the resulting o/p ratio of \hhco{}.

Finally, the current SMA observations are sensitive to the colder
outer disk, where \hhco{} is formed predominantly by grain surface
chemistry. Future observations at higher-angular resolution will allow
us to measure variations of the \hhco{} o/p ratio in the disk. A
direct comparison of the o/p ratio with the gas temperature structure
of the disk at different radii will help us test different formation
scenarios and elucidate what information can be extracted from o/p
measurements.






%

\section{Conclusions}
\label{sec:conclusions}

We have presented SMA observations (at $\sim3-7''$ angular
resolution) of 3 ortho and 2 para \hhco{} lines in the HD~163296
protoplanetary disk and constrained the disk-averaged \hhco{} o/p
ratio. We complemented these observations with one line of \phhco{} at
290~GHz at higher angular resolution ($\sim1''$) to determine the
\hhco{} abundance distribution. We consider a range of \hhco{}
distributions, locating \hhco{} at different heights in the disk, and
modeled the observed visibilities to constrain the best-fit o/p ratio
to the SMA observations. We derive an \hhco{} o/p value of $1.8-2.8$,
depending on the adopted \hhco{} spatial distribution, corresponding
to a spin temperature of 11-22~K. We also derive a rotational
temperature of 24~K from the flux ratio of the three ortho
lines. These results suggests that the bulk of \hhco{} forms by CO
hydrogenation on the surface of dust grains, in agreement with the
chemical model predictions, and is later released into the gas-phase
through some desorption mechanism.

Observations of o/p ratios could be powerful diagnostics of the
physical and chemical conditions of the gas. Moreover, future
resolved \hhco{} observations of disks, where regions with low o/p
ratios can be disentangled from regions presenting the statistical
value of 3, have great potential to constrain the formation pathway of
\hhco{} (gas-phase vs. grain surface chemistry) in disks.

\begin{acknowledgements}
V.V.G. and J.C acknowledge support from the National Aeronautics and
Space Administration under grant No. 15XRP15\_20140 issued through the
Exoplanets Research Progam.  This paper makes use of the following
ALMA data: ADS/JAO.ALMA\# 2012.1.00681.S. ALMA is a partnership of ESO
(representing its member states), NSF (USA) and NINS (Japan), together
with NRC (Canada), NSC andASIAA (Taiwan), and KASI (Republic of
Korea), in cooperation with the Republic of Chile. The Joint ALMA
Observatory is operated by ESO, AUI/NRAO and NAOJ.
\end{acknowledgements}

\end{document}